# In-memory photonic dot-product engine with electrically programmable weight banks

Wen Zhou,[1,†] Bowei Dong,[1,†] Nikolaos Farmakidis,[1] Xuan Li,[1] Nathan Youngblood,[1,2] Kairan Huang,[1] Yuhan He,[1] C. David Wright,[3] Wolfram H. P. Pernice,[4,5] and Harish Bhaskaran[1,*]

[1]Department of Materials, University of Oxford, Parks Road, OX1 3PH Oxford, United Kingdom
[2]Current Address: Department of Electrical and Computer Engineering, University of Pittsburgh, Pittsburgh, PA, 15261, USA
[3]Department of Engineering, University of Exeter, Exeter EX4 4QF, United Kingdom
[4]Institute of Physics, University of Münster, Heisenbergstr. 11, 48149 Münster, Germany
[5]Heidelberg University Kirchhoff-Institute for Physics Im Neuenheimer Feld 227, 69120 Heidelberg, Germany

[†]These authors contributed equally: W. Zhou and B. Dong
[*]Corresponding author: harish.bhaskaran@materials.ox.ac.uk

## Abstract

**Electronically reprogrammable photonic circuits based on phase-change chalcogenides present an avenue to resolve the von-Neumann bottleneck; however, implementation of such hybrid photonic–electronic processing has not achieved computational success. Here, we achieve this milestone by demonstrating an in-memory photonic–electronic dot-product engine, one that decouples electronic programming of phase-change materials (PCMs) and photonic computation. Specifically, we develop non-volatile electronically reprogrammable PCM memory cells with a record-high 4-bit weight encoding, the lowest energy consumption per unit modulation depth (1.7 nJ/dB) for Erase operation (crystallization), and a high switching contrast (158.5%) using non-resonant silicon-on-insulator waveguide microheater devices. This enables us to perform parallel multiplications for image processing with a superior contrast-to-noise ratio (≥ 87.36) that leads to an enhanced computing accuracy (standard deviation σ ≤ 0.007). An in-memory hybrid computing system is developed in hardware for convolutional processing for recognizing images from the MNIST database with inferencing accuracies of 86% and 87%.**

## Introduction

Non-von Neumann computing architectures, which physically collocate the data storage and analogue signal processing functionalities, are exceptionally suited to artificial intelligence (AI) applications and outperform digital processors especially in computer vision, achieving higher energy efficiency and fidelity.[1-3] Notably, in-memory computing hardware in the electronic domain based on transistor-programmed memristive crossbar arrays can be massively scaled to form '1T1R' type architectures,[4] which enables selective access and precise multi-level conductance programming of non-volatile memory elements in large-scale systems.[4-6] Recent interest has shifted towards in-memory photonic processing, where the clock frequency can be extended significantly beyond the GHz range.[7]

In-memory photonic computing based on photonic integrated circuits (PICs) represents a paradigm shift in harnessing parallelised data processing with appealing features such as ultrahigh clock frequency, massive parallelism, picosecond signal latency and ultrabroad bandwidth in optical signal processing.[7-9] Compared with digital electronic accelerators, in-memory photonic computing systems promise 1−3 orders of magnitude improvement in both compute density and energy efficiency,[7,10] which are critical in dealing with the heavy workload associated with running deep learning algorithms. Leveraging a silicon optoelectronic platform offers an opportunity for system scale-up based on low-cost, wafer-scale, and complementary metal–oxide–semiconductor (CMOS) compatible manufacturing.[11] Recent progress in implementing photonic neural networks (PNNs) is based on reconfigurable photonic–electronic hybrid PICs,[12,13] such as universal unitary networks and diffractive networks composed of Mach–Zehnder interferometers (MZIs),[14-16] broadcast-and-weight networks composed of microring resonators,[17] and deep optoelectronic networks.[18] These PNNs display key metrics including dense throughput and low latency using reliable electronic control of synapses with high-bit precision[19] and spiking neurons.[20] Photonic–electronic hybrid AI hardware implementations are more practical and competitive candidates than all-optical networks[21] for efficient *in situ* training[22] and self-calibration of weight banks[23] so as to suppress error accumulation in deep PNNs with good reproducibility.

In contrast to volatile optoelectronic PICs,[12] photonic non-von Neumann architectures leverage non-volatile waveguide memory cells[24] as functional layers integrated with silicon photonics to carry out computational tasks within the waveguide memory.[25] The key technology is developing a non-volatile photonic computing system using waveguide memory component that achieves multi-bit programming and thereby enables one to carry out in-memory computing with high precision. This has been done so far using phase change materials (PCMs), mostly of the absorbing variety, such as $Ge_2Sb_2Te_5$ (GST),[7,25,26] which is more reliable for photonic computing than those relying on precise phase control.[27,28] The change in effective refractive indices of these PCM cells can be achieved using optical or electrical pulses to induce a phase transition between the crystalline and amorphous states.[10] Optical programming of PCMs has been widely studied[24,26,29] and exploited for scalar and matrix-vector multiplication[7,25,30,31] due to high encoding precisions and reliable operations.[32] All-optical in-memory tensor cores have been recently developed with the capability of processing trillions of multiply-accumulate (MAC) operations per second using PCM-cladded $Si_3N_4$ waveguide crossbar arrays.[7] Due to the nonvolatility of the PCM cells, there is no additional energy consumption in weight holding, which is beneficial for convolutional operations with fixed kernels. For a small-scale silicon-on-insulator (SOI) tensor core, reconfigurable MZIs have been utilized for routing energetic optical pump pulses generated by a wall-plug or an integrated laser and an amplifier to optically program individual PCM cells.[30] Yet, as reconfigurable circuits, operation of these all-optical systems relies on two sequential steps: (i) optical programming for weight bank setting, (ii) optical signal probing for computing, i.e., separating the programming and computing. This complicates the system architecture and the inherent working mode, hindering its programming flexibility and scalability for a large-scale system. On the other hand, the mainstream SOI waveguide platform should be adopted for the CMOS-compatible optoelectronic integration with well-developed high-speed transceivers and mature ion-implantation technology. Considering all-optical programming on a SOI platform, it would be difficult to realize a large switching contrast and high-bit operation. The reported contrast is merely 15% with a small-area (a few $\mu m^2$) PCM switching and nearly 10 distinguishable memory levels.[33] To be more practical, it is conceivable to have a photonic processor that interfaces with electronic microcontrollers, and thus electrically programmable PCM cells are of paramount importance to set the multistate in a non-volatile manner for realizing the in-memory photonic computing.

To date, electrical pulse induced Joule heating has enabled the large-area phase switching of PCMs atop ion-implanted waveguide microheaters,[28,34-36] graphene microheaters,[37,38] ITO microheaters,[39] and plasmonic nanogap devices.[40,41] These device-level demonstrations have been exploited for optical switches[34,35] and tunable optical couplers.[36] Their functionality for in-memory photonic computing is, however, absent due to a limited number of encoding levels addressable ($\leq$ 8) in the non-resonant straight waveguide microheater devices,[28,34-41] which substantially limits the precision. Crucially, the following metrics arguably remain key and are required for advancement from a single PCM device to a computing system:

(i) multi-bit non-volatile electrical programming on the SOI waveguide platform
(ii) low energy consumption per unit modulation depth (unit: nJ/dB) for energy-efficient electrical programming
(iii) large switching contrast to enhance the contrast-to-noise ratio (CNR) and the computing accuracy
(iv) non-resonant and broadband operation of a PCM device for parallel computing

Here we report demonstration of an in-memory photonic–electronic computing system, which consists of high-performance computational memory cells based on the PCM-cladded microheaters with electrical control. Our PCM device and system resolve all the above critical issues, opening a promising avenue toward high-performance phase-change in-memory computing chips. The system features decoupled electronic programming of PCMs and optical probing of processed signals, which can be performed simultaneously. Specifically, the electronic circuits complete the weight bank setting and storage, and the optical circuits execute the scalar multiplication and MAC operation. This scenario could essentially combine the advantages of programming flexibility and scalability, high-speed optical signalling, and optoelectronic packaging synergies. Our electrically reprogrammable PCM cells achieve the following metrics: (i) a record-high 4-bit weight encoding, which outperforms 3-bit encoding of the previously developed SOI waveguide microheater devices,[28,34-37,39] (ii) the lowest energy consumption per unit modulation depth of 1.7 nJ/dB for the Erase operation (crystallization) of GST among waveguide microheater devices,[34-37,41] (iii) a very high switching contrast of 158.5% (modulation depth of 4.13 dB), which shows over ten times improvement compared with those using all-optical programming,[33] leading to enhanced CNRs ($\geq$ 87.36) and reduced computation error by 8.9 times achieved in our image processing experiment, (iv) parallel scalar multiplication using multiple wavelength division multiplexing (WDM) channels in image processing for brightness scaling, blurring and Sobel filtering. Lastly, a compact in-memory photonic–electronic computing system was electrically programmed to achieve edge detection for the implementation of convolutional neural networks (CNNs). Our experiments demonstrated high inferencing accuracies of 86% and 87% for recognizing images from the MNIST database, which compare favourably with previous PNNs.[15,16,21] We further estimated a compute density of 7.3 TOPS/mm$^2$ (Tera-operations per second per mm$^2$ chip area), a compute efficiency of 10.0 TOPS/W, and the energy consumption per MAC operation of 0.2 pJ/MAC at a data rate of 25 Gb/s and 16 WDM channels for an integrated in-memory photonic-electronic chip. Thus, our prototype system provides a viable path for the in-memory photonic–electronic computing with flexible programming, high-bit operation, low energy consumption, and high computational accuracy based on the SOI waveguide platform with prospective applications in intelligent edge devices for computer vision, speech recognition, autonomous driving, and signal processing.

**Results**

**In-memory photonic–electronic computing platform**

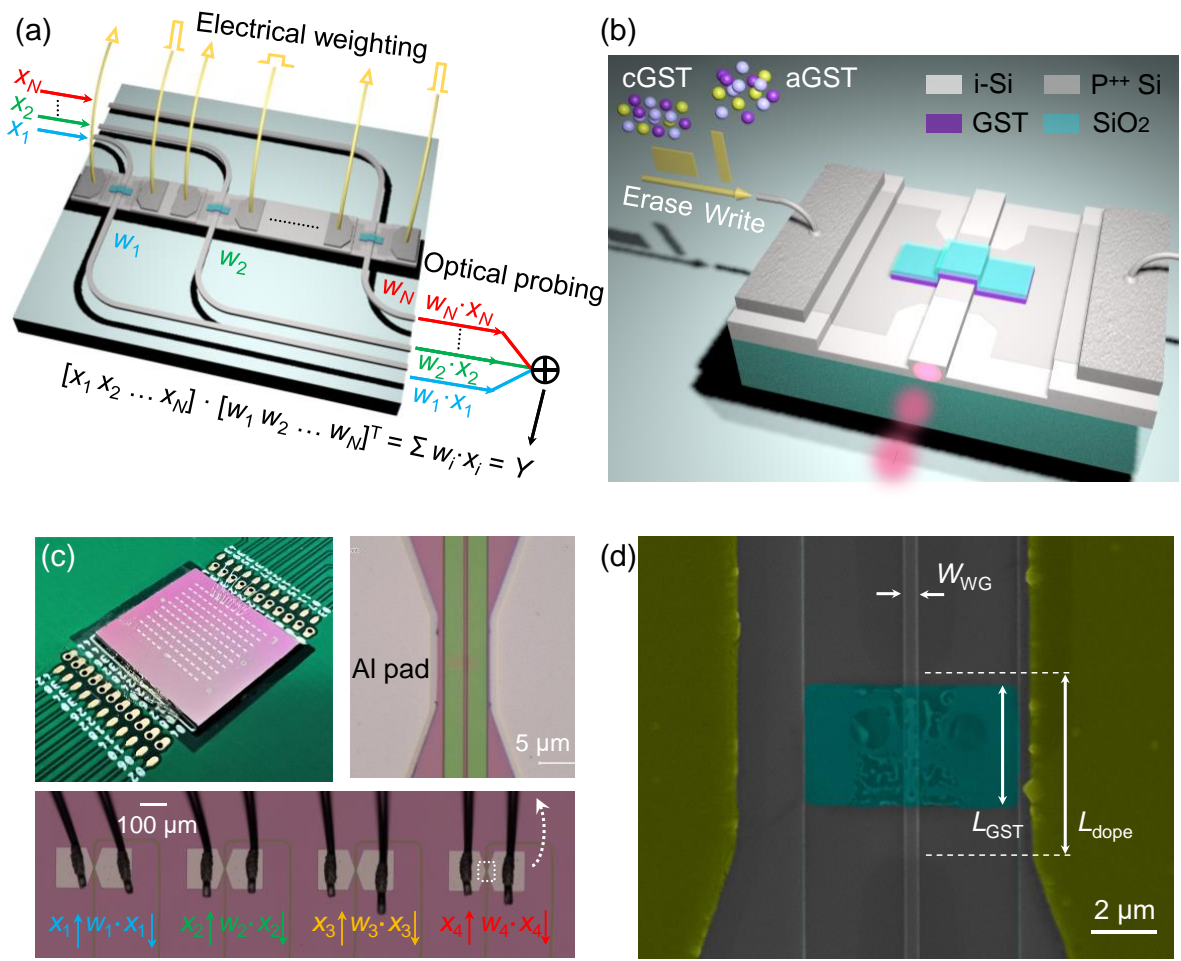

**Figure 1. An in-memory photonic–electronic computing platform.** (a) Conceptual illustration of an in-memory photonic–electronic dot-product engine and its working principles. (b) Schematic of an ion-implanted silicon-on-insulator (SOI) waveguide microheater cladded with a $Ge_2Sb_2Te_5$ (GST) thin film as a memory cell and a weight bank, simultaneously. c/aGST: crystallized/amorphized $Ge_2Sb_2Te_5$; i-Si: intrinsic silicon. (c) Top-view of a fabricated and wire bonded SOI photonic chip (1 cm × 1 cm) with a printed circuit board, and zoom-in optical micrographs of a small-scale in-memory photonic–electronic dot-product engine and a non-volatile waveguide memory cell. (d) A SEM image of an ion-implanted SOI waveguide microheater (measured doping length = 5.4 μm) with $GST/SiO_2$ (measured length = 3.5 μm and thickness = 30/50 nm) thin films (blue) and aluminum (Al) ground-signal (GS) electrodes (yellow).

Figure 1a illustrates a 3D schematic of an in-memory photonic–electronic dot-product engine consisting of an array of electrical programmable non-volatile GST memory cells, which act as reconfigurable weight banks [$w_1, w_2, \ldots w_N$] for dot-product operation. The input vector [$x_1, x_2, \ldots x_N$] is encoded in the amplitudes of probe light at different wavelengths ($\lambda_1, \lambda_2, \ldots \lambda_N$) generated by a broadband light source. Such probe energy is well below the power required to induce a phase transition in the GST cells. After encoding, the optical signal is transmitted through the GST cell and is subsequently and partially absorbed. Mathematically, it performs scalar multiplication with output amplitude of $w_i \cdot x_i$, in which the multiplicand $w_i$ is mapped to the transmittance of the $i$-th GST cell defined by its material state.[25] At the output ports of this dot-product engine, optical fields are combined incoherently in the optical domain to complete the summation of optical signals $\Sigma w_i \cdot x_i$ by using a $N$-to-1 wavelength division multiplexer (MUX), thus implementing a dot-product operation: [$x_1, x_2, \ldots x_N$]·[$w_1, w_2, \ldots w_N$]$^T$. Crucially, our proposed scalable in-memory photonic–electronic computing platform features decoupled electronic programming of weight banks and optical probing of weighted addition. As a critical component in this platform, a waveguide memory cell as shown in Fig. 1b is based on a boron-implanted SOI waveguide microheater (grey region) cladded with GST and $SiO_2$ thin films for non-volatile electrical programming. It is designed with a narrow doping strip covering a portion of a straight waveguide for efficient local Joule heating. Chip fabrication includes 120-nm shallow etching of the 220-nm-thick silicon device layer, heavily $P^{++}$ doping and aluminum (Al) pads (thickness = 300 nm) deposition as ground-signal (GS) electrodes for probe contact or wire bonding. Current–voltage (I–V) scanning shows a linear response due to Ohmic contact between Al pads and doped silicon (see Supplementary Fig. 1). GST and $SiO_2$ thin films (thickness = 30/50 nm) were last deposited on top of a doped SOI waveguide by radiofrequency (RF) magnetron sputtering. A 50-nm-thick $SiO_2$ capping was used to protect the GST thin film to avoid oxidation and delamination, thus enabling reliable operation (For detailed fabrication steps, see Methods and Supplementary Fig. 2). Top left of Fig. 1b shows our programming scheme using a single shot of short and high (long and low) amplitude electrical pulse for amorphization (crystallization) of a GST cell. As a result, transmission of the probe light can be modulated due to varied attenuation of the GST cell after programming. Figures 1c and 1d respectively show top-view of a fabricated and wire bonded SOI photonic chip (1 cm × 1 cm), an optical micrograph of a small-scale in-memory photonic–electronic dot-product engine, a zoom-in optical micrograph and a scanning electron microscopic (SEM) image of an electrically reconfigurable GST memory cell with a measured doping length ($L_{dope}$) of 5.4 µm and a total resistance of 238 Ω. Measured length of the GST/$SiO_2$ thin film ($L_{GST}$) is 3.5 µm (For detailed geometric parameters of our design, see Supplementary Fig. 3). There is an offset of patterned electrodes with respect to the waveguide due to overlay misalignment in electron-beam lithography. After programming by electrical pulses, large amorphization areas embedded in the crystalline matrix can be clearly observed in Fig. 1d due to the difference in conductivity between aGST and cGST. And the weight ($w_i$) is determined by the amorphous-to-crystalline ratio of the $i$-th GST cell, which can be electrically programmed by choosing proper electrical pulse parameters. Please note that only low-energy electrical pulses (< 9 nJ) were applied to avoid ablation of GST cells.[34,42]

## Multilevel GST cells for scalar multiplication

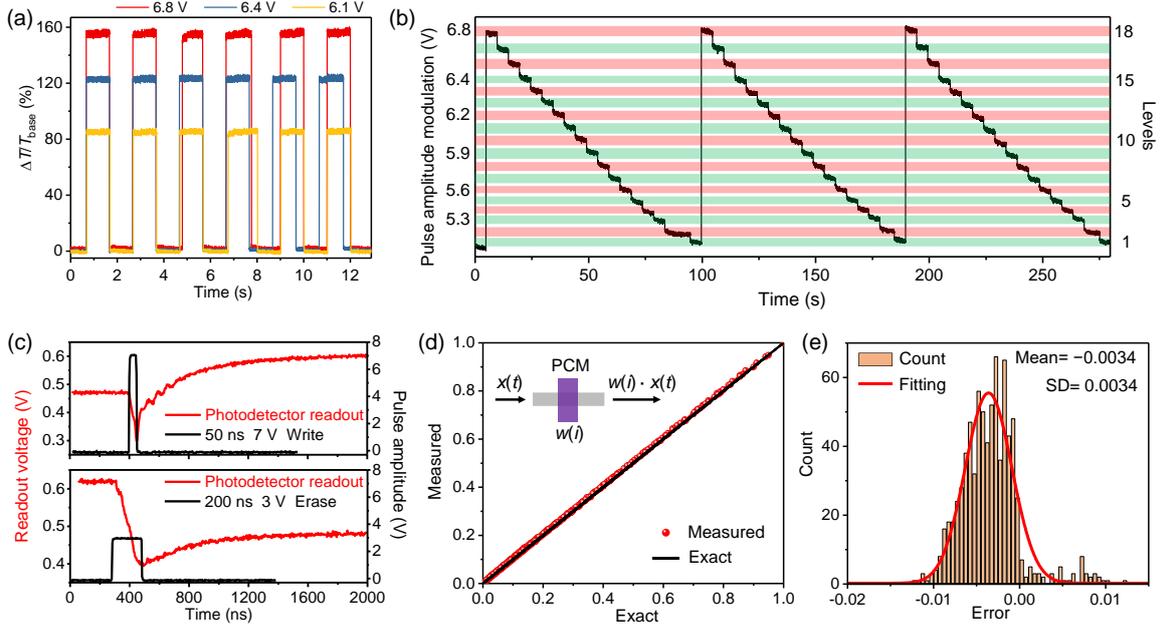

**Figure 2. Electrically reprogrammable photonic waveguide memory cells with over 4-bit encoding levels for scalar multiplication.** (a) Reversible electrical binary programming of a GST/SiO$_2$ cell ($L_{GST}$ = 2.5 μm, thickness = 30/50 nm and $L_{dope}$ = 3 μm) by 6.8V (red), 6.4V (blue), and 6.1 V (orange) 50-ns Write pulses and fixed 3V 200-ns Erase pulses. (b) Multilevel operation of a device showing over 4-bit non-volatile memory levels using pulse amplitude modulation (PAM). (c) Temporal switching dynamics (red curves) showing amorphization to a higher transmission level achieved with a single 50-ns 7V rectangular Write pulse (black curve), and recrystallization back to the baseline achieved by a single 200-ns 3V rectangular Erase pulse (black curve). (d) Measured scalar multiplication results versus exact results by performing 784 multiplication operations ($w(i) \times x(t)$) mapped by 16 different encoding levels ($w(i)$, $i$ = 1 to 16) of a GST cell as multiplicand (reached with write pulse energy between 5.2 nJ and 8.8 nJ), and 49 random input amplitudes of a probe light ($x(t)$, $t = (j-1)\cdot\Delta t$, $\Delta t$ = 1 ms, and $j$ = 1 to 49) as multiplier modulated sequentially by a variable optical attenuator (VOA) at a frequency of 1 kHz. (e) A histogram of computational error calculated by subtracting the measured scalar multiplication from the exact. The histogram is fitted by a Gaussian distribution (red solid curve). SD: standard deviation.

First, we demonstrated reversible binary and multilevel operations of GST cells based on an experimental setup elaborated in the Supplementary Fig. 4. Figure 2a shows the temporal optical waveform when a GST cell is switched back and forth upon sending electrical pulses. A single 50-ns-wide rectangular pulse with pulse amplitudes of 6.1 V (orange), 6.4 V (blue) and 6.8 V (red) partially amorphized the GST cell with incremental amorphous-to-crystalline ratios. A single 200-ns 3V rectangular pulse fully recrystallized the GST cell with transmission ($T$) back to the baseline ($T_{base}$). Considering the photodetection noise,[25] $T_{base}$ is the measured and averaged transmission baseline for a fully crystallized GST cell, and $T$ is measured and averaged transmission after a switching event for a partially amorphized GST cell. The switching contrast is defined as $\Delta T/T_{base}$, where change in transmission is defined as $\Delta T = T - T_{base}$. A switching contrast was measured as high as 158.5% for a microheater with a doping length of 3 µm and 2.5-µm-long GST/SiO$_2$ cell atop. Measured total resistance of the device is 261.5 Ω. Loss of 0.59 dB/µm was measured due to free carrier absorption in the heavily doped silicon waveguide (Supplementary Fig. 5). Hence, due to P$^{++}$ doping, the insertion loss of a 3-µm-long waveguide microheater is 1.78 dB. Based on a 3D finite element method (FEM) modelling, we found that the heating area expands with increasing doping length ($L_{dope} \leq 10$ µm) of the waveguide microheater, which can be used for switching a larger area of GST if required (Supplementary Fig. 6). However, we chose $L_{dope}$ of 3 µm to balance the trade-off between switching contrasts and optical losses. Binary operation can be used for optical switching applications as demonstrated by previous works.[34,35,37,41] However, multi-bit programming of waveguide memory cells is a basic requirement for photonic computing to obtain a high numerical accuracy. We further evaluated the performance of addressing deterministic multiple states by exploiting pulse amplitude modulation (PAM). A sequence of 17 Write pulses was sent with monotonically decreased amplitudes from 6.8 V to 5.2 V and a fixed pulse width of 50 ns followed by a recrystallization pulse with fixed amplitude of 3V and pulse width of 200 ns. PAM enables precise controlling of the amorphous-to-crystalline ratio of a GST cell. Figure 2b shows time traces of multilevel operation repeated by 3 cycles. 18 unique levels indicated by shaded green and red areas were resolved. Each level corresponds to a partially crystalline state, which was addressed with Write pulse energies between 5.2 and 8.8 nJ. And each 200-ns 3V Erase pulse consumes 6.9 nJ. The capacitive energy consumption is negligible.[43,44] Energy consumptions of 6.9 nJ (Erase) and 8.8 nJ (Write) are lower than those of reported microheater devices.[10] Such over 4-bit data encoding and storage capability may provide a leap forward for the in-memory photonic computing with electrical control. Figure 2c shows switching dynamics of the GST cell. Upon sending electrical pulses, readout voltage of a high-speed photodetector first drops due to thermo-optical effect.[24] At the end of the pulse heating, transmittance of the GST cell reaches the minimum and subsequently rises to another equilibrium state. The post-excitation dead times were measured to be 232 and 356 ns for Write and Erase, respectively. In terms of operational speed, it here requires 282 ns and 556 ns for Write and Erase, respectively. Supplementary Figure 7 also shows operation of our device with repetitive optical switching over 100 cycles and repetitive I–V scanning of microheater. Supplementary Table 2 compares performances of our experimentally demonstrated devices with the previously reported non-resonant straight waveguide GST cells with electrical controlling.[34-37,41] Our GST cells exhibit the highest encoding/storage levels reported so far and the lowest energy consumption per unit modulation depth in the Erase (recrystallization) process. With these advantages, our devices can be applied to in-memory photonic computing, instead of only optical switching.

Scalar multiplication is essential for neuromorphic computing due to heavy MAC workloads in PNNs.[14] We next demonstrated in-memory scalar multiplication operation of our GST memory cells with over 4-bit encoding precision. $T$ is the measured absolute transmittance of a GST cell with 16 discrete values between $T_{base}$ and $T_{max}$. $P_{in}$ between 0 and $P_{max}$ (= 0.35 mW) is the input power generated by a continuous wave (CW) probe laser after VOA modulation, which was sent down to a GST cell. We calibrated and normalized input power ($P_{in}/P_{max}$) of a VOA versus bias voltage and transmittance [$(T-T_{base})/(T_{max}-T_{base})$] of our GST cell versus Write pulse amplitude with a fixed pulse width of 50 ns (Supplementary Fig. 8) to obtain the exact $x(t) \times w(i)$ ($x(t)$ and $w(i) \in [0, 1]$). Transmittance of a GST cell versus

Write/Erase event needs to be calibrated independently due to fabrication error of microheater devices. In our testing, an electric Write pulse was first sent to a GST cell to reach a transmission level ($T$) followed by light probing, which was encoded sequentially with 49 random values of $P_{in}$ between 0 and 0.35 mW. We noted that $x(t) \times w(i)$ can be also expressed as: $(P_{in} \cdot T - P_{in} \cdot T_{base})/(P_{max} \cdot T_{max} - P_{max} \cdot T_{base})$, in which absolute transmission through the device including $P_{in} \cdot T$, $P_{in} \cdot T_{base}$, $P_{max} \cdot T_{max}$, $P_{max} \cdot T_{base}$ were recorded to calculate measured results of scalar multiplication in a post-processing step. A subtraction of $(P_{in} \cdot T - P_{in} \cdot T_{base})$ was performed on a digital computer to correct offset ($T_{base} \neq 0$). Instead, offset correction can be implemented by using the balanced photodetection method with estimated reduced energy consumption and signal processing time in the circuits (see Supplementary Note 10 for details). Accuracy of multiplication was then examined. Figure 2d shows a good matching between the exact and measured results of multiplication. The standard deviation (SD) of the residual error is as low as 0.0034, which is almost one order of magnitude lower than those reported in ref. 25, and the mean error is −0.0034 fitted by a Gaussian distribution as shown in Fig. 2e.

# In-memory parallel multiplication operations for image processing

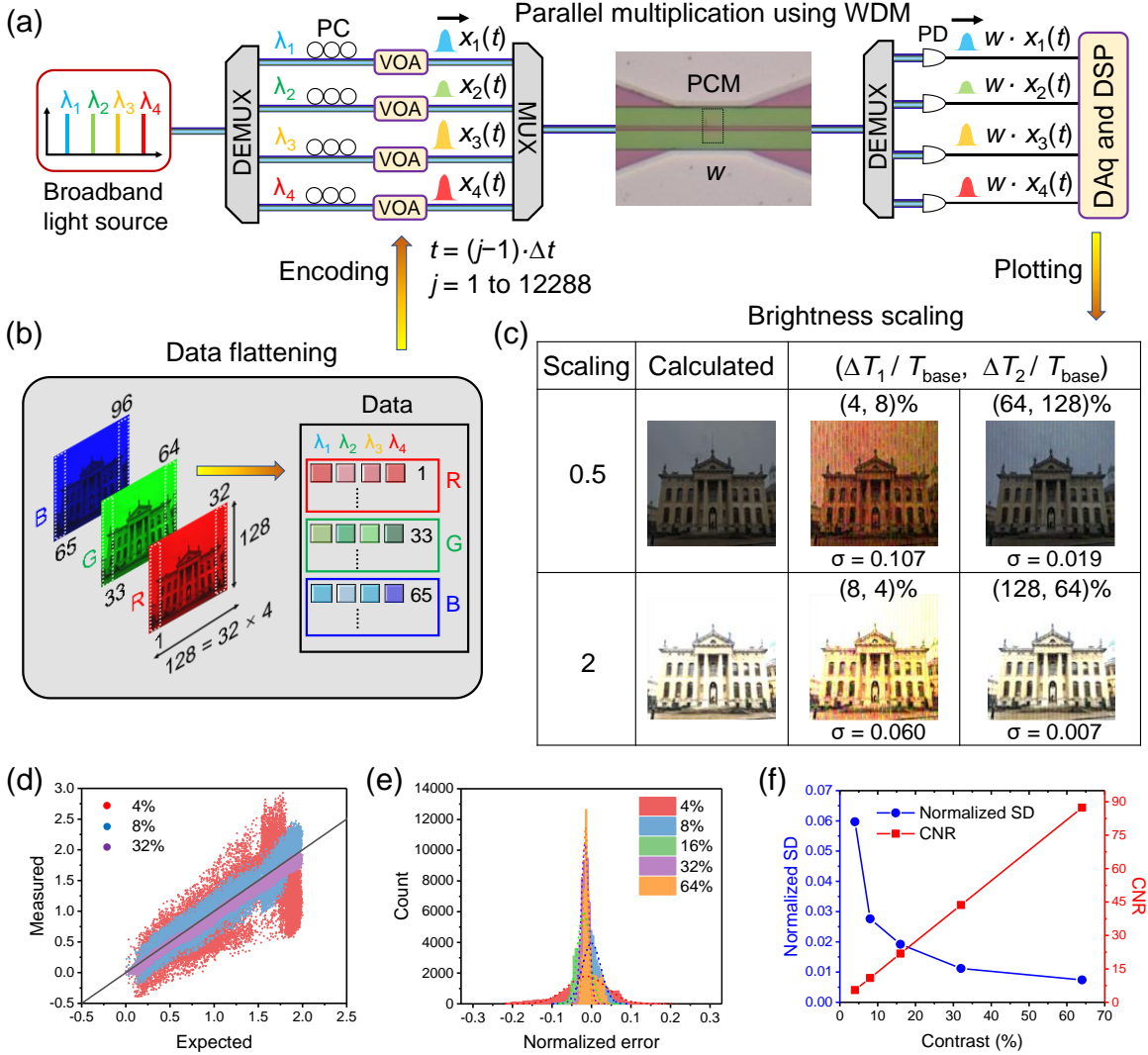

**Figure 3. WDM enabled parallel multiplication for image processing**. (a) An experimental setup for parallel multiplication by a GST cell. (b) Flattened pixel data of an RGB image encoded in amplitudes of four different wavelength channels by four VOAs. (c) Brightness scaling by performing light transceiving of all encoded pixels through a GST cell with varied switching contrast $\Delta T_{1(2)}/T_{base}$, followed by image plotting for visualization with labelled SD ($\sigma$) (The original image showing the Clarendon Building in the University of Oxford was photographed by the authors). (d)–(f) Examples of measured results for brightness scaling (scaling factor = 2) versus the exact results (d), histograms of normalized error fitted with Gaussian distributions (e), normalized SD and contrast-to-noise ratio (CNR), which is defined as $(T - T_{base})/SD(T)$ and $SD(T)$ is the standard deviation of the detected and normalized transmission mixing with photodetection noise, versus switching contrasts ($\Delta T_2/T_{base}$) of a GST cell (f).

One of the predominant advantages of photonic computing is parallel data processing enabled by the WDM scheme. We therefore study this property in image processing applications. Figure 3a shows an experimental setup for four-channel parallel multiplication as a proof-of-concept demonstration (see Methods). By this multiplexing, the processing time of an entire image can be reduced by a factor of four. As shown in Fig. 3b, an input image with a size of 128 × 128 showing the Clarendon Building at the University of Oxford (photographed by the authors) was decomposed into RGB channels, which were preprocessed to allow for rearranging into a 4 by 12288 (= 128 × 96) matrix for data flattening. In each timeframe of the computation, probe light at four separated wavelength channels ($\lambda_1$ to $\lambda_4$) generated by a broadband light source was encoded as one row of this flattened matrix using VOAs. These separated channels were multiplexed and combined to pass through a single GST cell. At the output of the device, channels were demultiplexed into four paths and detected by photodetectors simultaneously. Light in parallel channels was probed with nearly identical switching contrast after passing through the same GST cell upon electronic programming (Supplementary Fig. 10). Weight of the GST cell ($w$) was used to scale the amplitudes of signals ($x_1(t)$–$x_4(t)$, $t = (j-1)\cdot\Delta t$, $\Delta t = 1$ ms, and $j = 1$ to 12288). The scaling factor ($S$) is defined as $[(T_1-T_{base})/(T_2-T_{base})]$, where $T_{1(2)}$ is the transmittance of a GST cell after electrical programming. The scaling factor can be rewritten as $\Delta T_1/\Delta T_2$, and $\Delta T_1$ can be set as $0.5\cdot\Delta T_2$ or $2\cdot\Delta T_2$ to achieve brightness scaling with scaling factors of 0.5 or 2, respectively. We also note that $S \times x_i(t)$ ($i = 1$ to 4) can be expressed as: $[(T_1-T_{base})/(T_2-T_{base})]\cdot(P_{in}/P_{max}) = (P_{in}\cdot T_1 - P_{in}\cdot T_{base})/(P_{max}\cdot T_2 - P_{max}\cdot T_{base})$, in which absolute transmission through the device including $P_{in}\cdot T_1$, $P_{in}\cdot T_{base}$, $P_{max}\cdot T_2$, and $P_{max}\cdot T_{base}$ were recorded to calculate brightness scaling results in a post-processing step. In the experiment, the parallel multiplication process of an entire RGB image was repeated for various switching contrasts of a GST cell ($\Delta T/T_{base} = 0\%$, 4%, 8%, 16%, 32%, 64%, 128%) with multiplication results stored for post-processing and imaging plotting. As a direct comparison in Fig. 3c, we plotted images after applying brightness scaling (scaling factor = 0.5 or 2) using a GST cell with programmed ($\Delta T_1/T_{base}$, $\Delta T_2/T_{base}$) = (4%, 8%), (8%, 4%), (64%, 128%) or (128%, 64%). The measured images with higher contrasts of (64%, 128%) and (128%, 64%) match well with the theoretically calculated results, showing SDs of 0.019 and 0.007, respectively. However, the change of hue and noise are high in the measured images with contrasts of (4%, 8%) and (8%, 4%). Thus, a large $\Delta T_{1(2)}/T_{base}$, and therefore, a correspondingly high CNR, which is defined as $(T - T_{base})/SD(T)$, leads to a lower computational error. $SD(T)$ is the standard deviation of the normalized detected transmission, attributed mainly to photodetection noise. To provide quantitative analysis on the computational error of brightness scaling, Fig. 3d shows three sets of measured results versus the exact results at switching contrasts ($\Delta T_2/T_{base}$) of 4%, 8%, and 32% for doubling the brightness. Figure 3e shows histograms of computational error calculated by subtracting the measured results from the exact results. Histograms were further fitted by Gaussian distributions to extract SD as shown in Figure 3f. We observed a sharp decrease in SD versus switching contrast ($\Delta T_2/T_{base}$). SD is suppressed from 0.060 to 0.007 and CNR is enhanced from 5.46 to 87.36 by increasing $\Delta T_2/T_{base}$ from 4% to 64%. Errors are mainly from shot noise and thermal noise in photodetection (0.79%, 0.74%, 0.81%, and 1.07% for the four channels) with a 3-dB bandwidth of 11.6 kHz and light source power drift over time (1.82%, 3.59%, 2.89%, and 4.31% for the four channels) due to instability of our broadband light source (see Supplementary Note 12 for details). Boxcar averaging and error distributions of the three color tones indicate that the change of hue is not caused by noise, but mainly caused by laser power drift (see Supplementary Note 13 for details).

As a further example, we exploited parallel multiplication for realizing advanced image filtering (Supplementary Note 14). Similarly, it exhibits good matching between the measured and theoretical results from applying kernels of blurring and edge detection with large switching contrasts. In the image blurring application, SD is reduced from 0.071 to 0.008 by increasing $\Delta T_{step}/T_{base}$ from 4% to 64% (Supplementary Fig. 18). Thus, large switching contrasts achieved by electrical programming of PCMs using waveguide microheaters are crucial for image processing and feature extraction in CNNs based on the above comparison.

**An in-memory photonic–electronic dot-product engine for image recognition**

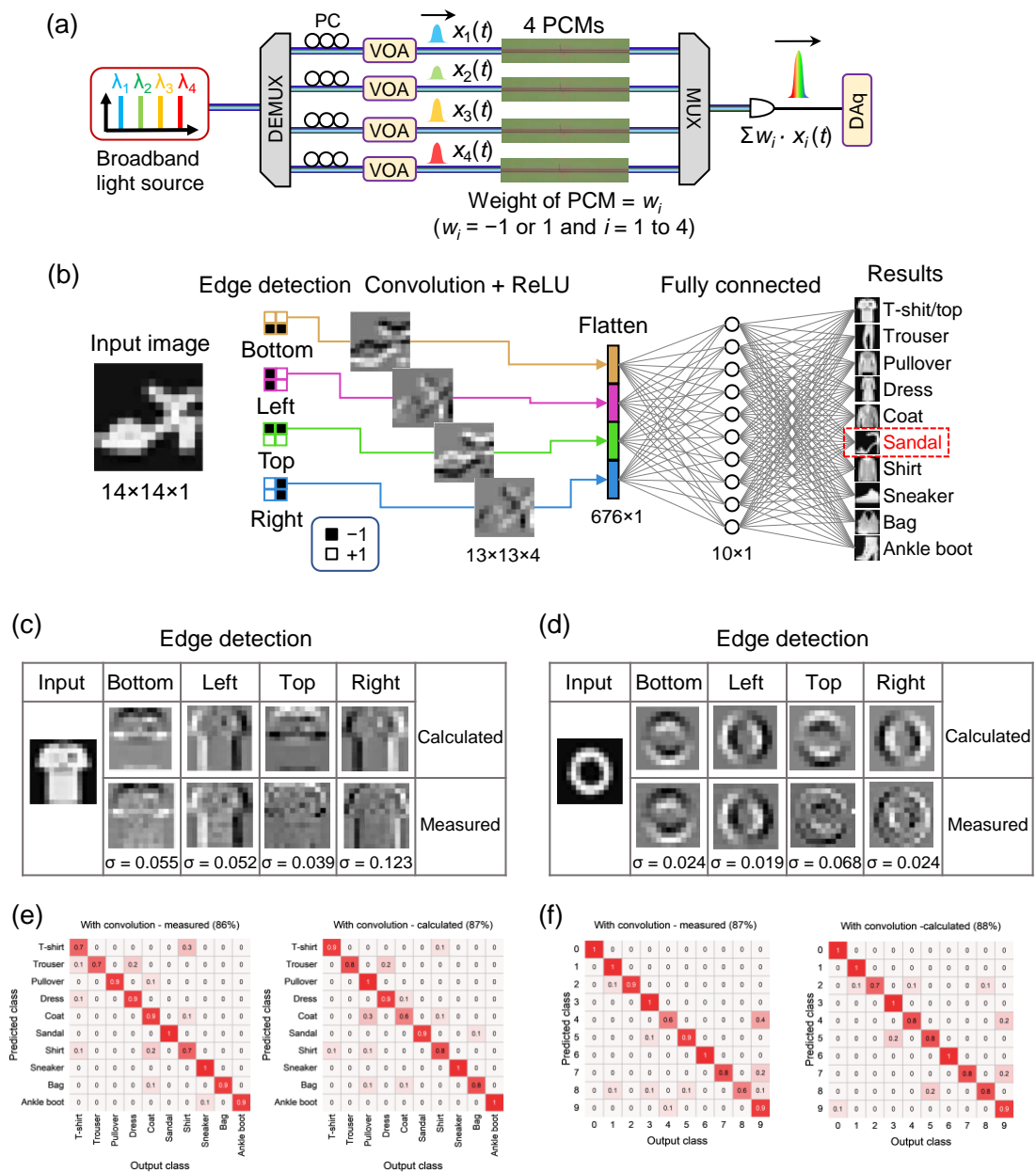

**Figure 4. Fashion product and digit recognition using CNNs.** (a) An experimental setup for implementing the dot-product operation. (b) A CNN architecture for both fashion product and handwritten digit recognition. Insets of fashion products are shown for illustration purpose. The photonic–electronic dot-product engine performs the convolution for MNIST fashion products[45] and handwritten digits[46] with an image size of 14 × 14. (c) and (d) Extracted features showing measured results with labelled SDs compared with those of theoretically calculated ones using a 32-bit computer for a T-shirt (c) and digit zero (d). (e) and (f) Confusion matrices showing comparable inferencing accuracy for fashion product recognition with 86% from experiment and 87% from calculation (e), and digit recognition with 87% from experiment and 88% from calculation (f).

Having demonstrated the parallel multiplication function for image processing, we next performed convolutional operations in the optical domain by exploiting a compact in-memory photonic–electronic system consisting of four GST cells, which represent a 2 × 2 kernel matrix and can be addressed with flexible electronic programming. As shown in Fig. 4a, four different wavelength channels were demultiplexed from a broadband light source, and were separately encoded by VOAs to represent a 2 × 2 patch of a 14 × 14 8-bit grey scale image in each timeframe of the computation. As a result, 169 (13 × 13 for valid padding) patches were generated and organized into a 4 × 169 matrix for data encoding in amplitudes of probe light. At outputs, probe signals were combined incoherently to obtain a time series of kernel-patch dot-product results ($\Sigma w_i \cdot x_i(t)$, $i$ = 1 to 4, $t = (j-1) \cdot \Delta t$, $\Delta t$ = 1 ms, and $j$ = 1 to 169). Please note that four scalar multiplications $w_1 \cdot x_1(t)$, $w_2 \cdot x_2(t)$, $w_3 \cdot x_3(t)$, and $w_4 \cdot x_4(t)$ were performed in parallel in the optical domain. And 169 repetitions of dot products of $\Sigma w_i \cdot x_i(t)$ were implemented sequentially by updating $x_i(t)$ ($i$ = 1 to 4). Thus, there are 169 × 4 MAC operations in total in the optical domain for convolving an image of 14 × 14 pixels. To map $w_i$ of ±1 onto a positive range of light transmittance, $w_i$ is defined as $2 \cdot (T_i - T_{ave})/\Delta T$ and $T_i \in [T_{min}, T_{max}]$, where $T_{ave} = (T_{min} + T_{max})/2$, $\Delta T = T_{max} - T_{min}$, and $T_{max}$ ($T_{min}$) is the maximum (minimum) transmittance of a GST cell by electrical programming. And the convolutional operation can be expressed as: $\Sigma [2(T_i - T_{ave})/\Delta T] \cdot (P_i/P_{max})$, where $P_i \in [0, P_{max}]$ ($i$ =1 to 4) and $x_i(t) = P_i/P_{max}$. With this mapping, 2-quadrant multiplication can be implemented and weights of GST cells represent bipolar signals by offsetting the background value of $T_{ave}$.

We benchmarked our dot-product engine with classification tasks on elementary datasets such as MNIST fashion product[45] and handwritten digits[46] using a CNN architecture as shown in Fig. 4b. As an example, ten-class fashion products were fed into the input layer of a CNN for optical signal encoding. The data were then convolved with four 2 × 2 kernels with elements of ±1 to generate four activation maps with an image size of 13 × 13 showing four detected edges. The computing accuracy and error distribution of the convolutional processing results are shown in Supplementary Fig. 20. The convolved data was then post-processed. After nonlinear activation by a rectified linear unit (ReLu) function, four activation maps were flattened into a 676 × 1 vector, which was then fed into a fully connected layer with ten neurons to output a final 10 × 1 vector for showing classification results. 400 images of fashion products/handwritten digits were convolved with four different kernels using our photonic–electronic dot-product engine, and their output data were used for training the 676 × 10 weight banks of a fully connected layer in our CNN by a digital computer. In this way, computational errors due to noise and drift were included in the training step to obtain a robust CNN. And another 100 images were fed into a trained CNN for generating a confusion matrix. To elaborate, edge detection results of grey-scale images with labelled SDs were examined with visualized examples of T-shirt and digit zero as shown in Figs. 4c and 4d, respectively. These measured results of highlighted edges (bright outlines) match well with those of theoretically calculated results. Inferencing accuracies for fashion product and digit (Figs. 4e and 4f) show good agreement between experimental measurement (86% and 87%) and theoretical calculation (87% and 88%). Supplementary Figure 21 shows detailed evolution of loss and accuracy versus epoch during neural network training for both Fashion-MNIST[45] and digit-MNIST[46] datasets. We noted that prediction accuracies were improved with the help of a convolutional layer compared with those of the fully connected neural networks without any convolutional layer. The computational load of the optical processing in the presented CNN is discussed in Supplementary Figure 22. And our CNN exhibits one of the highest prediction accuracies for the MNIST fashion product compared with those of the start-of-the-art PNNs previously reported[15,16,21] (see Supplementary Table 4 for details).

**Discussion**

In summary, we have demonstrated an in-memory photonic–electronic hybrid computing system based on the non-volatile GST memory cells, which support flexible electrical programming and high-throughput optical computing. Our electrically reprogrammable GST cells show record-high 4-bit encoding levels, the lowest energy consumption per unit modulation depth for the crystallization process (1.7 nJ/dB), and a high switching contrast (158.5%) using non-resonant SOI waveguide microheater devices. We then exploited parallel scalar multiplication using a GST cell with a very low error (SD = 0.007) and enhanced CNR for image processing. We have experimentally revealed the key role of enhancing the CNR in reducing the computation error by 8.9 times in image processing tasks. For neuromorphic computing, we implemented convolutional layers in CNNs based on an in-memory photonic–electronic dot-product engine with inferencing accuracies of 87% and 86% for MNIST handwritten digit and fashion product recognition, respectively, which were comparable with software calculation. In PNNs, a large switching contrast produces a wide variable range in summed energy to drive the cascaded artificial neuron crossing its threshold. An accumulated energy ($E_{min}$) estimated as 270.8 pJ, which is far below the switching threshold (420 pJ) of an integrate-and-fire artificial neuron,[47] can be reached based on $E_{min} = E_{max}/(\Delta T_{max}/T_{base}+1)$ when the contrast ($\Delta T_{max}/T_{base}$) is 1.585 and the maximally accumulated energy ($E_{max}$) is 700 pJ corresponding to the on-state of neuron. A large contrast enabled by electrical programming is a key advantage in the operation of PNNs with reliable cascading with neurons. Our results show that implementing the in-memory photonic–electronic hybrid platforms offer scalable and flexible programming, parallel optical signal processing, and CMOS-compatible wafer-scale fabrication. We believe that further advances may come from the following aspects: reducing insertion losses by optimizing design of ion implantation (e.g., using lightly doped microheaters), system scaling up using crossbar array networks[7] or resonator-based broadcast and weight networks with low insertion losses[17], and boosting data throughput by monolithic integrating the state-of-the-art silicon photonic modulators and germanium photodetectors, which may drive the system at tens of Gigabits per second (Gb/s) per WDM channel. Our presented dot-product engine can be scaled up to photonic tensor cores that implement matrix-matrix multiplication in a single clock, as described in Supplementary Note 16. Using the photonic tensor core, assuming a data rate of 25 Gb/s, 16 WDM channels for parallel computation and an estimated chip area of 56.02 mm$^2$, compute density of an in-memory photonic–electronic chip is estimated to be 7.3 TOPS/mm$^2$, which is significantly higher than those of current GPUs, tensor processing units (TPUs) and ASICs.[48-51] The compute efficiency is estimated to be 10.0 TOPS/W and the energy per MAC operation is 0.2 pJ/MAC (see Supplementary Note 16 for details).

## Methods:

### Device fabrication

In-memory photonic–electronic hybrid systems were fabricated on a SOI wafer (SOITECH) with a 220-nm-thick top silicon device layer and a 2-µm-thick buried oxide layer. The top silicon device layer was patterned by a JEOL JBX-5500 50kV electron-beam lithography (EBL) using a 400-nm-thick positive-tone resist (CSAR). The top silicon device layer was subsequently shallowly etched (etching depth = 120 nm) by inductively coupled plasma reactive ion etching (ICP-RIE, Oxford Instrument PlasmaPro) with SF6 and CHF3 gases, followed by $O_2$ plasma etching of CSAR resist. Next, a 3-µm-thick positive tone electron-beam resist poly(methyl methacrylate) (PMMA) 950 A8 was spin coated and patterned by EBL as ion implantation windows for the heavy boron doping. The ion implantation dosage was $3\times10^{15}$ ions/cm$^2$ at 10 KeV. After ion implantation, the SOI chips were immersed in acetone to remove the PMMA layer and the piranha solution to completely remove the resist residue. The SOI chips were then annealed at 950 °C for 3 mins in nitrogen atmosphere to activate the boron dopant by a Jipelec Jetfirst rapid thermal annealing processor. A 2-µm-thick double-layer PMMA (PMMA 495 A8 and PMMA 950 A8) were spin-coated on the chip surface, and the third step of EBL was performed to define the evaporation windows of aluminum (Al) pads on top of the doped silicon. Buffered oxide etching was performed to completely remove native oxide layer for achieving a good Ohmic contact between Al pads and doped silicon. A 300-nm-thick Al was subsequently deposited using thermal evaporation followed by lifting off. And the fourth step of EBL was performed to define the GST sputtering windows on top of the doped silicon waveguides. A stack of 30-nm/50-nm GST/SiO$_2$ was deposited using a magnetron sputtering system (PVD, AJA International Inc.) followed by a lift-off process. The GST and SiO$_2$ targets were respectively sputtered at 30 W RF power with 3 sccm Ar flow and 40 W RF power with 3 sccm Ar flow at a base pressure of $10^{-7}$ torr. Finally, the sputtered GST were lifted off followed by thermal annealing at 250 °C for 10 mins to fully crystallize the GST.

### Measurement setup

Optical measurement was performed using a fibre-chip coupling setup as illustrated in Supplementary Fig. 4. The input light generated from a continuous wave (CW) laser (7711A, Keysight Technologies) was aligned to an apodized waveguide grating coupler for light coupling under an off-vertical incident angle of 8°. The wavelength and power of the CW light were 1570.4 nm and 3.55 mW, respectively. Its polarization was optimized by a fibre polarization controller (Thorlabs, FPC032) to match the fundamental quasi-transverse-electric mode of the SOI waveguides. The input light power was controlled by a variable optical attenuator (Thorlabs V1550A) for data encoding at a modulation frequency of 1 kHz for demonstrating multiplication operation. Output light was spitted into two separated paths by a 1×2 fiber-optic coupler (TW1550R2A1) with a splitting ratio of 90:10. 10% of output light was collected by a low-noise photodetector (Newport 2011-FC) to obtain temporal trace of the GST cells upon phase transition for binary and multilevel operations. 90% of output light was collected by a high-speed photodetector (Newport 1811-FC) and a fast-sampling oscilloscope (TDS7404, Tektronix, Inc.) to characterize the thermo-optical response of the GST cells upon phase transition. In the electrical programming, RF electrical pulses were generated by a pulse generator (Tektronix AFG3102C) and DC electrical signals were generated by a source meter (Keithley 2614B), which were combined by using a bias tee (Mini-Circuits ZFBT-4R2GW+) and were applied to the Al pads through a ground-signal (GS) electrical probe (PicoProbe). DC current-voltage (I–V) sweeping was conducted to measure the total resistance of the waveguide microheater. To amorphize the GST cell, a single 50ns and 7V rectangular voltage pulse was sent. For crystallization, a single 200ns 3V rectangular voltage pulse was applied. To perform parallel multiplication, a broadband light source (SuperK COMPACT, NKT Photonics) was first filtered by a dual-channel passive filter (SuperK SPLIT, NKT Photonics) and was then demultiplexed by a $1 \times 16$ dense wavelength division multiplexing (DWDM) module to output multiple wavelength channels at the optical telecommunication wavelengths. These separated wavelength channels were modulated at a frequency of 1 kHz by VOAs (Thorlabs V1550A) to encode data of flattened pixels. After modulation, these channels were multiplexed again by another 1×16 DWDM module and were coupled into a single GST cell. At output port of the device, parallel channels were demultiplexed and detected by low-noise photodetectors (Newport 2011-FC). To perform convolutional operation for artificial CNNs, each wavelength channel demultiplexed from a broadband light source (SuperK COMPACT, NKT Photonics) was modulated at a frequency of 1 kHz by a VOA (Thorlabs V1550A) and a GST cell to perform scalar multiplication. At outputs, these wavelength channels were combined by another 1×16 DWDM module and detected by a photodetector (Newport 2011-FC). Data Acquisition and equipment controlling were performed by a microcontroller (National Instruments USB-6259 BNC, sampling rate of 1.25 MSamples/s) with analog outputs connecting to VOAs and analog inputs connecting with photodetectors.

**CNN model**

For both MNIST fashion product and handwritten digit recognition tasks, 500 images were taken from the MNIST database and compressed to 14×14 pixels. 80% of images (400 images) were used for training and 20% (100 images) for testing. The architecture of the employed CNN is shown in Fig. 4b. The input layer takes the pixel data (13×13 pixels for valid padding) and passes the data to a convolutional layer consisting of four pre-defined 2×2 kernels for edge detection, resulting in an output dimension of 13×13×4. The output is then flattened to 676×1 and activated by the rectified linear unit (ReLu) function. The activated 676×1 output is fed to a fully connected layer with 10 neurons, whose output is further converted to probabilities by a Softmax layer that shows the final classification results. The convolution layer was implemented using the in-memory photonic–electronic dot product engine. The subsequent ReLu function, fully connected layer, and Softmax function were constructed by software using Matlab R2021b Deep Learning toolbox. Weights of the fully connected layer were trained by Adam optimizer. The loss and accuracy versus epoch were monitored to ensure that the CNN was successfully trained to classify ten categories of images.

Acknowledgements

H. B., C. W., and W. P. acknowledge support from the EU H2020 programme (Grant No. 780848, Fun-COMP Project and Grant No. 101017237, PHOENICS Project).